\documentstyle[pra,aps]{revtex}

\title{Quantum version of the Monty Hall problem}
\author{A.P.\ Flitney\thanks{Email address: aflitney@physics.adelaide.edu.au},
D.\ Abbott\thanks{Email address: dabbott@eleceng.adelaide.edu.au}}
\address{Centre for Biomedical Engineering (CBME)
and Department of Electrical and Electronic Engineering, \\
Adelaide University, SA 5005, Australia}
\date{\today}

\begin{document}

\maketitle

\begin{abstract}
A version of the Monty Hall problem is presented where the players are
permitted to select quantum strategies.
If the initial state involves no entanglement
the Nash equilibrium in the quantum game offers the players
nothing more than can be obtained with a classical mixed strategy.
However, if the initial state involves
entanglement of the qutrits of the two players,
it is advantageous for one player
to have access to a quantum strategy while the other does not.
Where both players have access to quantum strategies
there is no Nash equilibrium in pure strategies,
however, there is a Nash equilibrium
in quantum mixed strategies
that gives the same average payoff as the classical game.
\end{abstract}


\section{INTRODUCTION}
Inspired by the work of von Neumann~\cite{neumann51}, classical information
theorists have been utilizing the study of games of chance since the 1950s.
Consequently, there has been a recent interest in recasting classical
game theory with quantum probability amplitudes, to create quantum games.
The seminal paper by Meyer in 1999~\cite{meyer99} pointed the way for
generalizing the classical theory of games to include quantum games. Quantum
strategies can exploit both quantum superposition~\cite{meyer99,goldenberg99}
and quantum entanglement~\cite{eisert99,benjamin00b}. There are many paradoxes
and unsolved problems associated with quantum information~\cite{god} and the
study of quantum game theory is a useful tool to explore this area.
Another motivation is that in the area of quantum communication, optimal
quantum eavesdropping can be treated as a strategic game with the goal of
extracting maximal information~\cite{brandt98}. It has also been suggested
that a quantum version of the Monty Hall problem may be of interest
in the study of quantum strategies of quantum measurement~\cite{li01a}.

The classical Monty Hall problem~\cite{savant91,gillman92} has raised much
interest because it is sharply counterintuitive. Also from an informational
viewpoint it illustrates the case where an apparent null operation does
indeed provide information about the system.

In the classical Monty Hall game the banker (``Alice'') secretly selects
one door of three behind which to place a prize. The player (``Bob'') picks a
door. Alice then opens a different door showing that the prize is not behind it.
Bob now has the option of sticking with his current selection or changing to the
untouched door. Classically, the optimum strategy for Bob is to alter his choice
of door and this, surprisingly, doubles his chance~\cite{savant91} of
winning the prize from $\frac{1}{3}$ to $\frac{2}{3}$.

\section{QUANTUM MONTY HALL}
A recent attempt at a quantum version of the Monty Hall problem~\cite{li01a}
is briefly described as follows: there is one quantum particle and
three boxes $| 0 \rangle$, $| 1 \rangle$, and $| 2 \rangle$. Alice selects a
superposition of boxes for her initial placement of the particle and Bob then selects
a particular box. The authors make this a fair game by introducing an additional
particle entangled with the original one and allowing Alice to make a quantum
measurement on this particle as a part of her strategy. If a suitable measurement
is taken after a box is opened it can have the result of changing the state of
the original particle in such a manner as to ``redistribute'' the particle
evenly between the other two boxes. In the original game Bob has a $\frac{2}{3}$
chance of picking the correct box by altering his choice but with this change
Bob has $\frac{1}{2}$ probability of being correct by either staying or
switching.

In the literature there are various explorations of quantum games
\cite{meyer99,eisert99,benjamin00b,li01a,marinatto00,benjamin00a,li01b,du00a,du00b,ng01,iqbal01,johnson01,flitney02}.
For example, the prisoner's dilemma~\cite{eisert99,benjamin00a,li01b},
penny flip\cite{meyer99}, the battle of the sexes~\cite{marinatto00,du00a}, and
others\cite{du00b,ng01,iqbal01,johnson01,flitney02}.
In this paper we take a different approach to Ref.~\cite{li01a}
and quantize the {\it original} Monty Hall game directly,
with no ancillary particles,
and allow the banker and/or player to access general quantum strategies.
Alice's and Bob's choices are represented by qutrits\cite{qubit}
and we suppose that they start in some initial state.
Their strategies are operators acting on their respective qutrit.
A third qutrit is used to represent the box ``opened'' by Alice.
That is, the the state of the system can be expressed as
\begin{equation}
| \psi \rangle = | o b a \rangle \;,
\end{equation}
where $a$ = Alice's choice of box,
$b$ = Bob's choice of box,
and $o$ = the box that has been opened.
The initial state of the system shall be designated as $| \psi_{i} \rangle$.
The final state of the system is
\begin{equation}
| \psi_{f} \rangle = ( \hat{\mbox{S}} \cos \gamma + \hat{N} \sin \gamma )
                 \, \hat{\mbox{O}} \, (\hat{I} \otimes \hat{B} \otimes \hat{A})
                        | \psi_{i} \rangle \;,
\end{equation}
where
$\hat{A} =$ Alice's choice operator or strategy,
$\hat{B} =$ Bob's initial choice operator or initial strategy,
$\hat{\mbox{O}} =$ the opening box operator,
$\hat{\mbox{S}} =$ Bob's switching operator,
$\hat{N} =$ Bob's not-switching operator,
$\hat{I} =$ the identity operator,
and $\gamma \in [0, \frac{\pi}{2}]$.
It is necessary for the initial state to contain a designation for an open box
but this should not be taken literally
(it does not make sense in the context of the game).
We shall assign the initial state of the open box to be $| 0 \rangle$.

The open box operator is a unitary operator that can be written as
\begin{equation}
\hat{\mbox{O}} = \sum_{ijk\ell} |\epsilon_{ijk}| \, | njk \rangle \langle \ell jk |
                \:+\: \sum_{j\ell} | mjj \rangle \langle \ell jj | \;,
\end{equation}
where
$|\epsilon_{ijk}| = 1$,
	if $i,j,k$ are all different and $0$ otherwise,
$m = (j + \ell + 1) (\mbox{mod} 3)$,
and $n = (i + \ell) (\mbox{mod} 3)$.

The second term applies to states where Alice would have a choice of box to open
and is one way of providing a unique algorithm for this choice\cite{Ooperator}.
Here and later the summations are all over the range $0, 1, 2$.
We should not consider $\hat{\mbox{O}}$ to be the literal action of
opening a box and inspecting its contents,
that would constitute a measurement,
but rather it is an operator that marks a box
(ie., sets the $o$ qutrit)
in such a way that it is anti-correlated with Alice's and Bob's choices.
The coherence of the system is maintained until the final stage
of determining the payoff.

Bob's switch box operator can be written as
\begin{equation}
\hat{\mbox{S}} = \sum_{ijk\ell} |\epsilon_{ij\ell}| \, | i\ell k \rangle
                \langle ijk |
		\: + \: \sum_{ij} | iij \rangle \langle iij | \;,
\end{equation}
where the second term is not relevant to the mechanics of the game
but is added to ensure unitarity of the operator.
Both $\hat{\mbox{O}}$ and $\hat{\mbox{S}}$ map each possible basis state to
a unique basis state.

$\hat{N}$ is the identity operator on the three-qutrit state.
The $\hat{A} = (a_{ij})$ and $\hat{B} = (b_{ij})$
operators can be selected by the players
to operate on their choice of box
(that has some initial value to be specified later)
and are restricted to members of SU(3).
Bob also selects the parameter $\gamma$ that controls the mixture of staying
or switching.

In the context of a quantum game
it is only the expectation value of the payoff that is relevant.
Bob wins if he picks the correct box, hence
\begin{equation}
\langle \$_{B} \rangle = \sum_{ij} |\langle ijj | \psi_{f} \rangle |^2 \;.
\end{equation}
Alice wins if Bob is incorrect, so
$\langle \$_{A} \rangle = 1 - \langle \$_{B} \rangle$.

\section{SOME RESULTS}
In quantum game theory
it is conventional to have an initial state $| 000 \rangle$
that is transformed by an entnaglement operator $\hat{J}$~\cite{eisert99}.
Instead we shall simply look at initial states
with and without entanglement.
Suppose the initial state of Alice's and Bob's choices
is an equal mixture of all possible states
with no entanglement:
\begin{eqnarray}
| \psi_{i} \rangle &=& | 0 \rangle \otimes
                \frac{1}{\sqrt{3}} (| 0 \rangle + | 1 \rangle + | 2 \rangle)
\otimes \frac{1}{\sqrt{3}} (| 0 \rangle + | 1 \rangle + | 2 \rangle) \;.
\end{eqnarray}
We can then compute
\begin{eqnarray}
\hat{\mbox{O}} (\hat{I} \otimes \hat{B} \otimes \hat{A}) | \psi \rangle
 &=& \frac{1}{3} \sum_{ijk} |\epsilon_{ijk}| \,
        (b_{0j} + b_{1j} + b_{2j}) (a_{0k} + a_{1k} + a_{2k})
        \, | ijk \rangle \\ \nonumber
 && \makebox[5mm]{} + \frac{1}{3} \sum_{j} (b_{0j} + b_{1j} + b_{2j})
(a_{0j} + a_{1j} + a_{2j}) \, | mjj \rangle \;; \\ \nonumber
\hat{\mbox{S}} \hat{\mbox{O}} (\hat{I} \otimes \hat{B} \otimes \hat{A})
        | \psi_{i} \rangle
 &=& \frac{1}{3} \sum_{ijk} |\epsilon_{ijk}| \, (b_{0j} + b_{1j} + b_{2j})
        (a_{0k} + a_{1k} + a_{2k}) \, | ikk \rangle \\ \nonumber
 && \makebox[5mm]{} + \frac{1}{3} \sum_{jk} \, |\epsilon_{jkm}| \,
        (b_{0j} + b_{1j} + b_{2j})
        (a_{0j} + a_{1j} + a_{2j}) \, | mkj \rangle \;,
\end{eqnarray}
where $m = (j+1) (\mbox{mod} 3)$.
This gives
\begin{eqnarray}
\label{pay_noent}
\langle \$_{B} \rangle &=&
\frac{1}{9} \cos^2 \gamma \sum_{jk} (1-\delta_{jk}) \,
        |b_{0j} + b_{1j} + b_{2j}|^2
 |a_{0k} + a_{1k} + a_{2k}|^2 \\ \nonumber
 && \makebox[5mm]{} + \frac{1}{9} \sin^2 \gamma \sum_{j}
        |b_{0j} + b_{1j} + b_{2j}|^2 |a_{0j} + a_{1j} + a_{2j}|^2 \;.
\end{eqnarray}

We are now in a position to consider some simple cases.
If Alice chooses to apply the identity operator,
which is equivalent to her choosing a mixed classical strategy where
each of the boxes is chosen with equal probability,
Bob's payoff is
\begin{equation}
\label{pay_Aident}
\langle \$_{B} \rangle =
        \left( \frac{2}{9} \cos^2 \gamma + \frac{1}{9} \sin^2 \gamma \right)
         \sum_{j} |b_{0j} +b_{1j} + b_{2j}|^2 \;.
\end{equation}
Unitarity of $B$ implies that
\begin{eqnarray}
\sum_{k} |b_{ik}|^2 &=& 1
	\makebox[1cm]{} \mbox{for} \; i = 0,1,2, \\ \nonumber
\makebox[1cm]{and} \sum_{k} b_{ik}^{*} b_{jk} &=& 0
        \makebox[1cm]{} \mbox{for}\; i,j = 0,1,2 \; \mbox{with} \; i \ne j \;,
\end{eqnarray}
which means that the sum in Eq.\ (\ref{pay_Aident})
is identically 3.
Thus,
\begin{equation}
\label{pay_cl}
\langle \$_{B} \rangle = \frac{2}{3} \cos^2 \gamma + \frac{1}{3} \sin^2 \gamma \;,
\end{equation}
which is the same as a classical mixed strategy where Bob chooses to switch
with a probability of $\cos^2 \gamma$ (payoff $\frac{2}{3}$)
and not to switch with probability $\sin^2 \gamma$ (payoff $\frac{1}{3}$).

The situation is not changed where Alice uses a quantum strategy
and Bob is restricted to applying the identity operator
(leaving his choice as an equal superposition of the three possible boxes).
Then Bob's payoff becomes
\begin{equation}
\langle \$_{B} \rangle =
        \left( \frac{2}{9} \cos^2 \gamma + \frac{1}{9} \sin^2 \gamma \right)
         \sum_{j}|a_{0j} +a_{1j} + a_{2j}|^2 \;,
\end{equation}
which, using the unitarity of $A$, gives the same result as Eq.\ (\ref{pay_cl}).

If both players have access to quantum strategies,
Alice can restrict Bob to at most $\langle \$_{B} \rangle = \frac{2}{3}$
by choosing $\hat{A} = \hat{I}$,
while Bob can ensure an average payoff of at least $\frac{2}{3}$
by choosing $\hat{B} = \hat{I}$ and $\gamma = 0$ (switch).
Thus this is the Nash equilibrium of the quantum game
and it gives the same results as the classical game.
The Nash equilibrium is not unique.
Bob can also choose either of
\begin{equation}
\label{Ms}
\hat{M}_1 = \left( \begin{array}{ccc}
                0 & 1 & 0 \\
                0 & 0 & 1 \\
                1 & 0 & 0
        \end{array} \right)
\makebox[1cm]{or}
\hat{M}_2 = \left( \begin{array}{ccc}
                0 & 0 & 1 \\
                1 & 0 & 0 \\
                0 & 1 & 0
        \end{array} \right) \;,
\end{equation}
which amount to a shuffling of Bob's choice,
and then switch boxes.

It should not be surprising that the quantum strategies produced nothing new
in the previous case
since there was no entanglement
in the initial state\cite{unentangled}.
A more interesting situation to consider is an initial state
with maximal entanglement between Alice's and Bob's choices:
\begin{equation}
| \psi_{i} \rangle = | 0 \rangle \otimes \frac{1}{\sqrt{3}}
        (| 00 \rangle + | 11 \rangle + | 22 \rangle) \;.
\end{equation}
Now
\begin{eqnarray}
\hat{\mbox{O}} (\hat{I} \otimes \hat{B} \otimes \hat{A}) | \psi_{i} \rangle
 &=& \frac{1}{\sqrt{3}} \sum_{ijk\ell} |\epsilon_{ijk}|
        \, b_{\ell j} a_{\ell k} \, | ijk \rangle
 + \frac{1}{\sqrt{3}} \sum_{j\ell}
        b_{\ell j} a_{\ell j} \, | mjj \rangle \;; \\ \nonumber
\hat{\mbox{S}} \hat{\mbox{O}} (\hat{I} \otimes \hat{B} \otimes \hat{A}) | \psi_{i} \rangle
 &=& \frac{1}{\sqrt{3}} \sum_{ijk\ell} |\epsilon_{ijk}| \,
        b_{\ell j} a_{\ell k} \, | ikk \rangle
 + \frac{1}{\sqrt{3}} \sum_{jk\ell} \, |\epsilon_{jkm}| \,
                b_{\ell j} a_{\ell j} \, | mkj \rangle \;,
\end{eqnarray}
where again $m = (j+1) (\mbox{mod} 3)$.
This results in
\begin{eqnarray}
\langle \$_{B} \rangle &=& \frac{1}{3} \sin^2 \gamma \sum_{j}
                |b_{0j} a_{0j} + b_{1j} a_{1j} + b_{2j} a_{2j}|^2 \\ \nonumber
 &&     + \frac{1}{3} \cos^2 \gamma \sum_{jk}
                (1-\delta_{jk}) \, |b_{0j} a_{0k} + b_{1j} a_{1k} +
                         b_{2j} a_{2k}|^2 \;.
\end{eqnarray}

First consider the case where Bob is limited to a classical
mixed strategy.
For example, setting $\hat{B} = \hat{I}$
is equivalent to the classical strategy of selecting
any of the three boxes with equal probability.
Bob's payoff is then
\begin{eqnarray}
\langle \$_B \rangle &=& \frac{1}{3} \sin^2 \gamma \,
        (|a_{00}|^2 + |a_{11}|^2 + |a_{22}|^2) \\ \nonumber
&& \makebox[5mm]{} + \frac{1}{3} \cos^2 \gamma \,
	(|a_{01}|^2 + |a_{02}|^2 + |a_{10}|^2
		+ |a_{12}|^2 + |a_{20}|^2 + |a_{21}|^2) \;.
\end{eqnarray}
Alice can then make the game fair by selecting an operator whose
diagonal elements all have an absolute value of $\frac{1}{\sqrt{2}}$
and whose off-diagonal elements all have absolute value $\frac{1}{2}$.
One such SU(3) operator is
\begin{equation}
\hat{H}
 = \left( \begin{array}{ccc}
        \frac{1}{\sqrt{2}} & \frac{1}{2} & \frac{1}{2} \\
        -\frac{1}{2} & \frac{3 - i \sqrt{7}}{4 \sqrt{2}} &
                \frac{1 + i \sqrt{7}}{4 \sqrt{2}} \\
        \frac{-1 - i \sqrt{7}}{4 \sqrt{2}} &
                \frac{-3 + i \sqrt{7}}{8} &
                \frac{5 + i \sqrt{7}}{8}
\end{array} \right) \;.
\end{equation}
This yields a payoff to both players of $\frac{1}{2}$,
whether Bob chooses to switch or not.

The situation where Alice is limited to the identity operator
(or any other classical strategy) is uninteresting.
Bob can achieve a payoff of 1 by setting $\hat{B} = \hat{I}$
and then not switching.
The correlation between Alice's and Bob's choice of boxes remains,
so Bob is assured of winning.
Bob also wins if he applies $\hat{M}_1$ or $\hat{M}_2$
and then switches.

As noted by Benjamin and Hayden\cite{benjamin00a},
for a maximally entangled initial state
in a symmetric quantum game,
every quantum strategy has a counterstrategy
since for any $U \in$ SU(3),
\begin{equation}
(\hat{U} \otimes \hat{I}) \frac{1}{\sqrt{3}}
        ( | 00 \rangle + | 11 \rangle + | 22 \rangle )
 = (\hat{I} \otimes \hat{U}^T) \frac{1}{\sqrt{3}}
        ( | 00 \rangle + | 11 \rangle + | 22 \rangle ) \;.
\end{equation}
Since the initial choices of the players are symmetric,
for any strategy $\hat{A}$ chosen by Alice,
Bob has the counter $\hat{A}^*$:
\begin{eqnarray}
(\hat{A}^* \otimes \hat{A}) \frac{1}{\sqrt{3}}
        ( | 00 \rangle + | 11 \rangle + | 22 \rangle )
 &=& (\hat{I} \otimes \hat{A}^{\dagger} \hat{A} ) \frac{1}{\sqrt{3}}
        ( | 00 \rangle + | 11 \rangle + | 22 \rangle ) \\ \nonumber
 &=& \frac{1}{\sqrt{3}} ( | 00 \rangle + | 11 \rangle + | 22 \rangle ) \;.
\end{eqnarray}
The correlation between Alice's and Bob's choices remains,
so Bob can achieve a unit payoff by not switching boxes.

Similarly for any strategy $\hat{B}$ chosen by Bob,
Alice can ensure a win by countering with $\hat{A} = \hat{B}^*$
if Bob has chosen $\gamma=0$,
while a $\gamma=1$ strategy is defeated by
$\hat{B}^* \hat{M}$,
where $\hat{M}$ is $\hat{M}_1$ or $\hat{M}_2$ given in Eq.\ (\ref{Ms}).
As a result there is no Nash equilibrium amongst pure quantum strategies.
Note that Alice can also play a fair game,
irrespective of the value of $\gamma$,
by choosing $\hat{B}^* \hat{H}$,
giving an expected payoff of $\frac{1}{2}$
to both players.
A Nash equilibrium amongst mixed quantum strategies can be found.
Where both players choose to play $\hat{I}$, $\hat{M_1}$ or $\hat{M_2}$
with equal probabilities neither player can gain an advantage over the classical
payoffs. If Bob chooses to switch all the time, when he has selected the same
operator as Alice, he loses, but the other two times out of three he wins.
Not switching produces the complementary payoff of $\langle \$_B \rangle =
\frac{1}{3}$, so the situation is analogous to the classical game.

\section{CONCLUSION}
For the Monty Hall game where both participants have access to quantum strategies,
maximal entanglement of the initial states produces the same payoffs as the
classical game.
That is, for the Nash equilibrium strategy the player, Bob, wins two-thirds
of the time by switching boxes.
If the banker, Alice, has access to a quantum strategy while
Bob does not, the game is fair, since Alice can adopt a strategy with an expected
payoff of $\frac{1}{2}$ for each person,
while if Bob has access to a quantum strategy
and Alice does not he can win all the time.
Without entanglement the quantum game confirms our expectations by offering
nothing more than a classical mixed strategy.

\section*{ACKNOWLEDGMENTS}
This work was supported by GTECH Corporation Australia
with the assistance of the SA Lotteries Commission (Australia).
Useful discussions with David Meyer, USCD, and Wanli Li, Princeton University,
are gratefully acknowledged.


\begin{references}
\bibitem{qubit} Qutrit is the three state generalization of the term qubit
which refers to a two-state system.
\bibitem{Ooperator} Note that this operator gives results for the opened box
that are inconsistent with the rules of the game if $\ell = 1$ or $\ell = 2$.
It is necessary to write the operator this way to ensure unitarity,
however, we are only interested in states where the initial value of the
opened box is $| 0 \rangle$, i.e., $\ell = 0$.
\bibitem{unentangled} Eisert\cite{eisert99} finds that with an unentangled
initial state, in quantum prisoners' dilemma, the players' quantum strategies
are equivalent to classical mixed strategies.
\end{references}
\end{document}